\documentclass[11pt] {IEEEtran}
\usepackage{amssymb,amsmath,graphicx,textcomp, graphicx, indentfirst, float, cite} 
\usepackage{graphicx}
\graphicspath{{images/}}
\usepackage{caption}
\usepackage{subcaption}
\expandafter\let\csname equation*\endcsname\relax
\expandafter\let\csname endequation*\endcsname\relax
\usepackage{mathtools}
\usepackage{braket}
\usepackage[utf8]{inputenc}
\usepackage[T1]{fontenc}
\usepackage{mathptmx}
\usepackage[markup=underlined]{changes}
\usepackage{hyperref}
\usepackage{cleveref}
\usepackage{etoolbox}
\usepackage{multirow}
\usepackage{booktabs,soul}
\graphicspath{ {images/} }

\begin{document}
\author{
\IEEEauthorblockN{
Hashir Kuniyil\IEEEauthorrefmark{1},
Asad Ali\IEEEauthorrefmark{1},
Syed M. Arslan\IEEEauthorrefmark{1},
Muhammad Talha Rahim\IEEEauthorrefmark{1},
Artur Czerwinski\IEEEauthorrefmark{2},
Saif Al Kuwari\IEEEauthorrefmark{1}
}

\IEEEauthorblockA{\IEEEauthorrefmark{1}
Qatar Center for Quantum Computing, College of Science and Engineering,
Hamad Bin Khalifa University, Qatar\\
{Corresponding Author:} hkuniyil@hbku.edu.qa
}

\IEEEauthorblockA{\IEEEauthorrefmark{2}
STARTOVA UMK Sp. z o.o., ul. Gagarina 7, 87-100 Torun, Poland}}

\title{Recovery-Induced Erasure Attack on QKD Systems}

\maketitle
\begin{abstract}
Detector dead time is typically treated as a fixed parameter in quantum key distribution (QKD) security analyses. In practice, however, the effective recovery time of single-photon avalanche photodiodes (SPADs) depends on the incident count rate. In this work, we demonstrate that this count-rate–dependent recovery nonlinearity constitutes a distinct attack primitive. We experimentally characterize the dead-time shift of a free-running SPAD under controlled broadband loading and observe a substantial increase in effective recovery time as the detected rate rises into the high photon count regime.

We show that recovery-induced availability reduction can be modeled as an adversarial erasure channel and derive a conservative bound on the signal detection probability under loading. Unlike previously studied detector-control or efficiency-mismatch attacks, the proposed mechanism does not rely on deterministic blinding or timing discrimination. Instead, count-rate–dependent recovery asymmetry induces basis-dependent suppression of detection probabilities ($p_\perp<p_\parallel$), converting mismatch-induced errors into loss. Particularly, we show in active-basis BBM92 systems, this effect reduces the observed quantum bit error rate (QBER) below the abort threshold while increasing erasure probability. Using experimentally measured detector recovery data, we quantify the parameter regime in which such stealth suppression is achievable.

These results establish count-rate–dependent detector recovery as a security-relevant vulnerability and show that countermeasures designed for timing-based efficiency mismatch do not directly address recovery-induced erasure (RIE) attack. Our findings underscore the need to incorporate detector recovery dynamics explicitly into practical QKD security models.
\end{abstract}

\section{Introduction}
Today's quantum technology allows the transmission of secret cryptographic keys over 400 km in optical fiber \cite{boaron2018secure} and supports other architectures, including long-range satellite communication \cite{yin2017satellite} and free-space ground links \cite{schmitt2007experimental}. Also, QKD systems now demonstrate high robustness against adverse environmental challenges \cite{wang2018practical, lasota2023robustness, seabrook2024surpassing}, leading many countries to adopt QKD preemptively against quantum computing threats \cite{lewis2022secure, lu2022micius}.\\
Although theoretically secure, many loopholes in the practical system concern its use in important applications. Historically, there have been multiple efforts to identify loopholes in QKD systems and develop strategies to eliminate them \cite{arslan2025device}. Prominent quantum attacks are detection efficiency mismatch (DEM) attack \cite{makarov2006effects, zhao2008quantum, fung2008security, jain2011device,rau2014spatial, chaiwongkhot2019eavesdropper, ma2005practical, gottesman2004security}, detector blinding attack \cite{sauge2011controlling,gerhardt2011full,lydersen2010hacking,fung2008security, Lydersen2010}, number splitting attack \cite{lutkenhaus2000security}. Many countermeasures have been proposed to counter these attacks \cite{hwang2003quantum,wang2005beating, lo2005decoy, gobby2004quantum, rau2014spatial, sajeed2015security, nauerth2009information}.

In this paper, we propose a quantum key distribution (QKD) attack strategy that exploits a previously unmodeled physical nonlinearity
of single-photon avalanche photodiodes (SPADs): the dependence of effective detector dead time on the incident count rate~\cite{stipvcevic2017advanced, krause2025exponential, meng2016ingaas,Gerrits2012, Zhang2009, liang2011low}. Specifically, we demonstrate how an eavesdropper (Eve) can exploit this recovery nonlinearity to convert detection events that would otherwise manifest as quantum bit errors into channel losses, thereby suppressing the observed QBER below the protocol's abort threshold while evading detection. This mechanism builds on the implicit assumption in standard QKD protocols that channel loss is adversarially benign and does not indicate eavesdropping~\cite{bennett1984proceedings,
bennett1992quantum}. In conventional security analyses, Eve's intercept-resend activity increases the QBER; the RIE attack subverts this relationship by converting the error signature into loss.

Prior work on detector-side attacks in QKD has addressed related but mechanistically distinct vulnerabilities. Bright-light blinding attacks force SPADs into linear operation via sustained
high-intensity illumination, enabling deterministic control of detector clicks with near-zero QBER at the cost of requiring detectable optical power levels~\cite{Lydersen2010, gerhardt2011full, sauge2011controlling}. Detection efficiency mismatch (DEM) and time-shift attacks exploit pre-existing asymmetries in the detector response curves across different measurement bases~\cite{makarov2006effects, zhao2008quantum, fung2008security}, achieving QBER suppression when the efficiency ratio between two detectors exceeds a critical threshold. The after-gate attack exploits residual sensitivity during the SPAD reset window to generate classical-regime responses, but requires precise timing relative to the gate boundary~\cite{weier2011quantum}. Importantly, all of these approaches either require deterministic, full control over detector state (as in blinding), or exploit static pre-existing hardware asymmetries (as in DEM). None exploits the dynamic, count-rate-dependent nature
of SPAD recovery.

The only prior attack to explicitly employ dead time as its principal mechanism is the dead-time attack of Weier et al.~\cite{weier2011quantum}, in which dim polarized blinding pulses force three of four detectors in a passive-basis-choice receiver into dead time, leaving only one active detector and deterministically controlling Bob's measurement outcome. This attack, however, relies
on (i) a fixed, binary dead-time model in which recovery is instantaneous once the dead-time interval $t_d$ elapses; (ii) a passive four-detector basis-selection architecture; and (iii) full deterministic suppression of detector availability. More recently,
Burenkov et al.~\cite{burenkov2010security} showed theoretically that at high transmission rates (pulse rate $\gtrsim 1/t_d$), intercept-resend naturally induces basis-dependent detection efficiency as a secondary effect of dead-time loading, but this analysis also assumes a fixed, binary recovery model and does not frame count-rate-dependent dead-time variation as an attack
primitive. A very recent contribution, the ``muted attack''~\cite{su2025security}, exploits dead-time loading in high-speed sinusoidal-gated SPADs with width discriminators, again under a fixed dead-time assumption and targeting a qualitatively different architectural setting.

In contrast, the present work demonstrates that the shift of the effective dead time itself -its dependence on the incident count rate constitutes a distinct and previously unanalyzed attack primitive. This count-rate-dependent recovery nonlinearity has been rigorously characterized in the detector physics literature~\cite{stipvcevic2017advanced, krause2025exponential}: the detection efficiency $\eta(t)$ recovers non-instantaneously and non-linearly following each avalanche, with the effective recovery time increasing substantially as the photon flux rises. The entire QKD security literature, however, models dead time as a step
function - binary availability with instantaneous recovery - and therefore this physical effect has not previously been incorporated into adversarial models. A recent defense-side paper~\cite{grasselli2025quantum} has begun to address basis-dependent detection probability as a security concern, but does not consider the adversarial engineering of such asymmetry through recovery-state manipulation.

The RIE attack exploits this gap through a two-stage intercept-resend architecture combining a polarization-structured pre-pulse with a subsequent resent signal pulse. The pre-pulse is prepared in Eve's measurement basis and carries the opposite bit value from the resent signal; its function is to place a specific detector at a controlled point on its sub-nominal recovery curve, creating a differential detection suppression $p_\perp < p_\parallel$ that selectively converts mismatch-induced error events into erasures. Crucially, this mechanism (i) does not require full detector blinding, (ii) does not require Eve to know Bob's basis choice in
real time, (iii) operates without deterministic timing control under a conservative non-deterministic pre-pulse model, and (iv) relies on no pre-existing hardware asymmetry. To our knowledge, the explicit use of count-rate-dependent dead-time variation to engineer an
adversarial erasure channel that suppresses the observed QBER below the abort threshold has not been previously analyzed.

This paper provides a rigorous analysis of the RIE attack and its security implications. We formalize Eve's strategy as an adversarial erasure channel, derive the conditions under which basis-dependent erasure asymmetry can be achieved below the QBER abort threshold, and
develop an information-theoretic framework quantifying Eve's mutual information advantage over the legitimate parties. Experimentally, we
characterize the count-rate-dependent dead-time shift of a free-running SPAD (Excelitas SPCM-AQRH-14-FC), confirming that the effective recovery time increases from approximately 23.3\,ns to
31.5\,ns at high count rates --- validating the physical foundation of the attack. We further evaluate the parameter regime in which stealth QBER suppression is achievable under the conservative, non-deterministic pre-pulse model, demonstrating the attack's feasibility even with restricted adversarial capability. Finally, we
discuss the distinction between RIE and efficiency-mismatch countermeasures, and identify mitigation strategies specifically tailored to recovery-induced erasure mechanisms.
This paper is organized as follows. In Section \ref{EveStrategy}, we present Eve’s attack strategy under the assumption that the detector dead time shifts in the high count-rate regime, illustrated using an active BBM92 setup. In Section \ref{tailoring_attack}, we model the detector recovery mechanism to realize the RIE attack and establish the corresponding security threshold. In section \ref{Mutual_information}, we derive the mutual information analysis assuming Eve is positioned between the legitimate parties. this section also gives complete details of shared information between all three parties in the attacking strategy. In Section \ref{exp}, we present experimental measurements of the dead time under varying count rates and demonstrate the feasibility of the attack under a non-deterministic pre-pulse model. Finally, in Section \ref{discussion}, we discuss potential mitigation strategies for recovery-induced erasure attacks.
\section{Eve's attacking scheme}\label{EveStrategy}
In this section, we describe a mechanism that induces basis-dependent erasure (i.e., events that do not contribute to the key) in an active-basis polarization receiver without requiring Eve to target a specific basis arm or to know Bob’s basis choice in real time. The method relies on count-rate–dependent detector recovery and a polarization-structured pre-pulse.

\begin{figure*}[t]
    \centering
    \includegraphics[width=\textwidth]{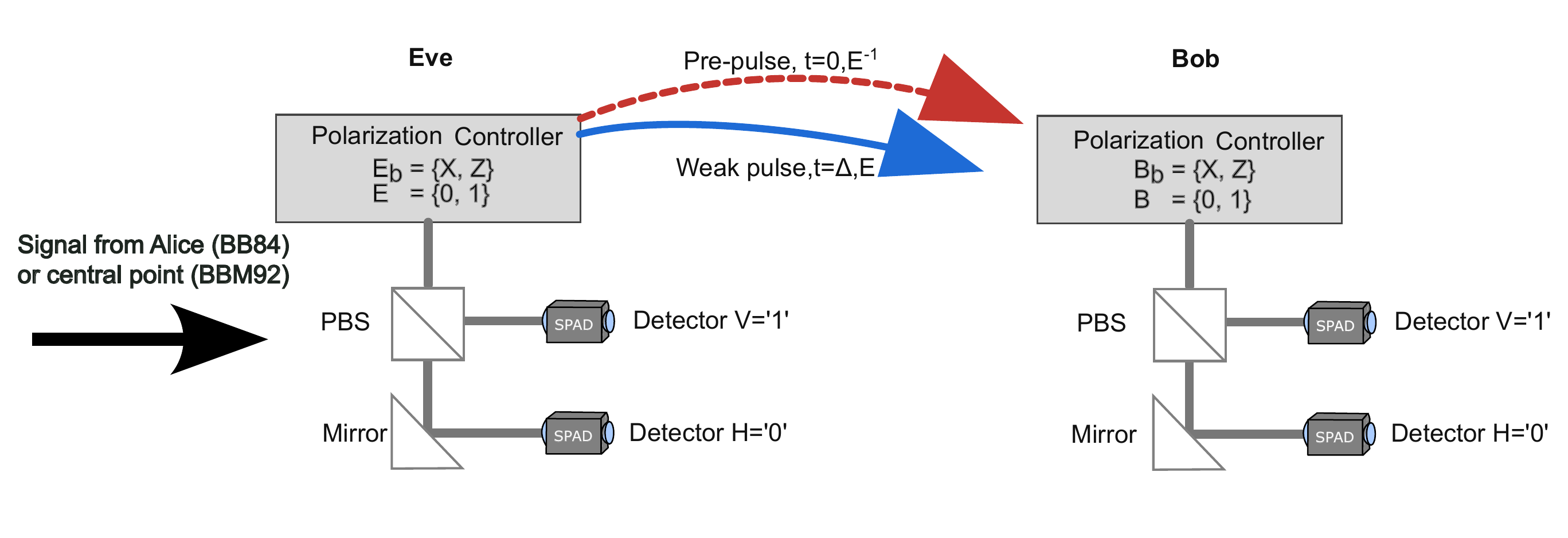}
    \caption{Schematic of the RIE attack, where Eve employs an apparatus identical to Bob’s. Based on the probabilistic measurement, Eve directs a combination of a pre-pulse and a weak pulse according to the attacking strategy.}
    \label{fig1}
\end{figure*}

We consider an active polarization receiver in which Bob selects the measurement basis using an electro-optic modulator (EOM) or a wave plate, followed by a single polarization beam splitter (PBS) and two single-photon detectors (this configuration is commonly used in BBM92 and BB84 protocol implementations). The same detector pair is used for both $Z$ and $X$ basis measurements; basis selection is implemented through polarization rotation prior to the PBS (see Fig.~\ref{fig1}) 

Let the computational basis $Z = \{\ket{H}, \ket{V}\}$ and the diagonal basis $X = \{\ket{D}, \ket{A}\}$, where
\begin{equation}
    \ket{D} = \frac{\ket{H}+\ket{V}}{\sqrt{2}}, \quad \ket{A} = \frac{\ket{H}-\ket{V}}{\sqrt{2}}.
\end{equation}
We assume the detectors exhibit a count-rate–dependent dead time $t_d(\lambda)$, where $\lambda$ is the observed count rate. A strong optical pre-pulse can temporarily reduce detector availability by increasing the effective dead time. This behavior stems from the slow recovery of the bias voltage in the SPAD's quenching circuit, where high count rates lead to cumulative charge accumulation and thermal effects that extend the time required to reset the detector to its operational state.

In BBM92/BB84 protocol, for each intercepted signal, Eve performs standard intercept–resend: She measures the incoming photon in basis 
$E_b \in \{Z, X\}$. She obtains bit value $E\in\{0,1\}$. She re-sends a weak signal pulse prepared in state $(E_b, E)$. To induce selective erasure, Eve additionally transmits a strong pre-pulse immediately before the resent signal pulse, with the following structure:
\begin{itemize}
    \item The pre-pulse is prepared in the same basis $E_b$.
    \item The pre-pulse carries the opposite bit value $\bar{E}$.
\end{itemize}
The resent signal pulse is delayed by a fixed interval $\Delta$ after the pre-pulse such that the detector loaded by the pre-pulse remains in dead time when the signal arrives.
If Bob measures in the same basis as Eve, the PBS routes polarization eigenstates deterministically: the pre-pulse, prepared in $(E_b, \bar{E})$, is directed to the detector corresponding to bit $\bar{E}$. The resent signal pulse, prepared in $(E_b, E)$, is directed to the detector corresponding to bit $E$. Because the two detectors are independent, the pre-pulse loads only the detector associated with $\bar{E}$, while the detector associated with $E$ remains available. Consequently, the signal photon experiences minimal suppression. We denote the click probability in this condition as $p_{\parallel}$, which remains close to the baseline detection probability.

If Bob measures in the basis orthogonal to Eve’s resent signal, the situation changes both qualitatively and quantitatively. A polarization eigenstate in one basis decomposes into an equal superposition in the other basis. such as 
\begin{equation}
    \ket{H} = \frac{\ket{D}+\ket{A}}{\sqrt{2}}, \quad \ket{V} = \frac{\ket{D}-\ket{A}}{\sqrt{2}}.
\end{equation}
Thus, when Bob measures in the orthogonal basis:

\begin{itemize}
    \item The pre-pulse state is split approximately equally between both detectors.
    \item The resent signal state is also split approximately equally between both detectors.
\end{itemize}

As a result, the pre-pulse increases the likelihood that both detectors are in dead time during the arrival of the signal photon. This reduces the effective click probability in the orthogonal-basis measurement.
\begin{table*}[h]
    \centering
    \caption{Click and error probabilities across all attack branches for a fixed Alice state ($Z0$). The table illustrates how Eve’s pre-pulse strategy targets specific states to suppress errors across different basis scenarios.}
    \label{tab1}
    \begin{tabular}{c c c c c c c c c}
        \hline
        Eve basis & Eve bit & Pre-pulse sent & Bob basis & Signal detector loaded & Click prob & Kept & Error & Effect \\
        \hline
        Z & 0 & Z1 & Z & No & $p_\parallel$ & Yes & No & Correct bit preserved \\
        Z & 0 & Z1 & X & Both partially & $p_\perp$ & No & --- & Irrelevant (discarded) \\
        X & 0 & X1 & X & No & $p_\parallel$ & No & --- & Irrelevant (discarded) \\
        X & 0 & X1 & Z & Both partially & $p_\perp$ & Yes & 50\% if clicks & Error suppression \\
       X & 1 & X0 & X & No & $p_\parallel$ & No & --- & Irrelevant (discarded) \\
        X & 1 & X0 & Z & Both partially & $p_\perp$ & Yes & 50\% if clicks &  Error suppression \\
        \hline
    \end{tabular}
\end{table*}
We denote this suppressed click probability as 
$p_\perp$, with Eve's target is to achieve,
\begin{equation}
    p_\perp<p_\parallel.
\end{equation}
Importantly, this asymmetry arises without Eve having knowledge of Bob’s basis choice. The selectivity emerges from the transformation properties of polarization states under basis rotation. This scheme is summarized in Table~\ref{tab1} when Alice sent in $Z$ basis and bit "$0$" ($Z0$) (Z1 and X-basis states are symmetric).

The above mechanism creates two distinct conditional detection probabilities:
\begin{align}
    p_\parallel &= P(\text{click} \mid B_b = E_b) \\
    p_\perp &= P(\text{click} \mid B_b \neq E_b)
\end{align}
In the intercept–resend scenario, the error-prone events correspond to those in which Bob measures in Alice’s basis while Eve measured in the orthogonal basis. These events fall into the $p_\perp$ category. By reducing $p_\perp$ relative to $p_\parallel$, Eve converts a fraction of potential bit errors into erasures (no-click events). Following this mechanism, the observed quantum bit error rate (QBER) on the sifted key becomes (we assume equal prior probabilities for Eve choosing $Z$ vs. $X$ basis),
\begin{equation}
    e_{obs} =\frac{p_\perp}{2(p_\perp+p_\parallel)}
    =\frac{r}{2(1+r)}.
    \label{eq4}
\end{equation}
Therefore, sufficient asymmetry $r = p_\perp/p_\parallel <<1$ suppresses the QBER below the protocol abort threshold ($r_{th}$) while increasing loss. This constitutes an adversarial erasure channel induced by detector recovery dynamics rather than by full detector blinding.
\section{rate-dependent detector dead time for tailoring the attack} \label{tailoring_attack} 

Let $t_d(\lambda)$ be the effective dead time as a function of observed count rate $\lambda$. The detector recorded arrival rate is defined
\begin{equation}
    \beta = \frac{\lambda}{1-t_d\lambda}
\end{equation}
where $\beta$ is true photon arrival rate in a detector. This is valid when $\lambda t_d<1$.
In a QKD round, Bob expects a signal photon in a narrow temporal window.

We define:
\begin{itemize}
    \item $A$: the event that the detector is available (not in dead time) at the signal arrival time.
    \item $p_0$: the conditional probability the signal causes a click if the detector is available (this includes quantum efficiency, coupling, basis optics, etc.).
\end{itemize}

Then, the signal click probability is:
\begin{equation}
    p_{click} = p_0 Pr(A)
\end{equation}
Assuming noise detections follow a Poisson process with observed rate $\lambda$, and neglecting correlations induced by dead time, the probability that the detector is available at the signal arrival time equals the probability that no registered detection occurred within the previous dead-time interval \cite{kuniyil2025optimizing} 
\begin{equation}
    Pr(A) \approx exp(-\lambda t_d(\lambda)). 
\end{equation}
Consequently, the signal click probability becomes
\begin{equation}
    p_{click} = p_0\space exp(-\lambda t_d(\lambda)),
\end{equation}
which models the transition from an ideal detection channel to an erasure channel due to rate-dependent detector dead time.

After sifting in protocols such as BB84/BBM92, Bob’s outcome per round is: a bit $B\in\{0,1\}$ if a click occurs, or an erasure $\varnothing$ otherwise. The effective channel is therefore a binary erasure channel (BEC) plus some bit-flip noise. Formally: with probability $\epsilon$ Bob outputs $\varnothing$ while with probability $1 - \epsilon$ Bob outputs a bit through a BSC with error $e$. Here,
$\epsilon = 1 - p_{\text{click}}(\lambda)$.
So the channel is: BEC($\epsilon$) + BSC($e$) (called a “erasure-and-error” channel). Based on this, we can model $p_\parallel$ and $p_\perp$ as
\begin{equation}
    p_{\parallel} = p_0 \exp\big(-\lambda_{\parallel} \, t_d(\lambda_{\parallel})\big),
    \label{eq11}
\end{equation}
\begin{equation}
    p_{\perp} = p_0 \, \exp\bigl(-\lambda_{\perp} \, t_d(\lambda_{\perp})\bigr),
    \label{eq12}
\end{equation}
where we assumed that all detectors have identical basic characteristics, $p_0$. Then the sifted key probability is
\begin{equation}
    P_{sift} = \frac{1}{2}\left(\frac{1}{2}p_\parallel+ \frac{1}{2}p_\perp \right)=\frac{p_\parallel+p_\perp}{4}
\end{equation}
In BBM92 protocol, if QBER satisfies $e_{obs}<0.11$ it assumes communication is safe. In our analysis, if Eve wants to hide its presence she has to achieve, from Eq.~(\ref{eq4}), 
\begin{equation}
    \frac{p_\perp}{p_\parallel}=r<0.282.
\end{equation}

\section{Information leakage in recovery induced erasure attack}\label{Mutual_information}
Mutual information between two parties can be generally quantified as
\begin{equation}
    I(X;Y) = H(X) - H(X|Y),
\end{equation}
where $H(X)$ is the Shannon entropy of binary random variable $X$, given by $H(x) = -x\log_2(x) - (1-x)\log_2(1-x)$, and $H(X|Y)$ is the conditional entropy of $X$ given $Y$. We will now derive mutual information formulas for our scheme, i.e., the mutual information between the legitimate parties Alice and Bob, $I(A;B)$, and the mutual information between Alice and Eve, $I(A;E)$. Throughout the paper, we assume binary encoding by Alice with equal probabilities, i.e. $P(A = 0) = P(A=1) = 1/2 $. In the QKD scheme, Bob's measurement outcomes are $B \in \{0, 1, \varnothing\}$. We assume that with probability $\epsilon_B$: $B = \varnothing$. Therefore, with probability $1-\epsilon_B$, Bob obtains a bit through a binary symmetric channel with crossover probability (QBER) $e_B$, i.e.
\begin{equation}
    P(B\neq A|B\neq\varnothing) = e_B.
\end{equation}
In this setting, when Bob obtained $B = \varnothing$, he learned nothing about Alice's key, therefore, the entropy reaches 
\begin{equation}
    H(A|B = \varnothing) = 1,
\end{equation}
when $B\in\{0,1\}$, it's outcome will have an error $e_B$, then,
\begin{equation}
    H(A|B \neq\varnothing) = h_2(e_B).
\end{equation}
Thus, on average
\begin{align}
H(A\mid B)
  &= P(B = \varnothing)\cdot 1 + P(B \neq \varnothing)\cdot h_2(e_B) \nonumber\\
  &= \epsilon_B + (1-\epsilon_B) h_2(e_B)
\end{align}
The mutual information between Alice and Bob in our scheme is
\begin{align}
    I(A;B) &= 1 - \bigl(\epsilon_B + (1-\epsilon_B)h_2(e_B)\bigr) \nonumber\\ 
           &= (1-\epsilon_B)\bigl(1 + h_2(e_B)\bigr)
\end{align}
The mutual information is scaled down by a factor of $1-\epsilon_B$, and non-erased keys behaves line BSC.
If we assume Eve's measurment outcomes are $E \in \{0, 1, \varnothing$\}, similar to Bob, Eve's probability outcomes with erasure probability $\epsilon_E$ and conditional error $1-\epsilon_E$ of non-erased events are
\begin{equation}
    P(E=\varnothing) = \epsilon_E, \hspace{0.5cm} P(E\neq A|E\neq\varnothing) = e_E.
\end{equation}
Finally, the mutual information between Alice and Eve can be derived as 
\begin{align}
    I(A;E) = (1-\epsilon_E)\bigl(1 + h_2(e_E)\bigr)
    \label{eq21}
\end{align}
In our setting, what matters is information per sifted detected bit (i.e., conditioned on Bob producing a sifted click). In that case, Bob has no erasure in the alphabet anymore, but Eve’s correlation with Alice becomes a mixture of two event classes:
\begin{itemize}
    \item Class M (match): Eve’s basis matched Alice’s $\rightarrow$ Eve knows the bit perfectly.
    \item Class U (unmatch): Eve’s basis mismatched Alice’s $\rightarrow$ Eve’s bit is independent of Alice (effectively random).
\end{itemize} 
Let the probability that a sifted detected event came from these classes be:
\begin{equation}
    \omega_M = P(M \mid \text{sifted}), \qquad \omega_U = 1 - \omega_M
\end{equation}
In class M, Eve has complete information about Alice bits, therefore, $H(A|E, M) = 0$. In class U, however, Eve has random information about Alice's keys. Thus, conditional information becomes $H(A|E, U) = 1$. Thus average $H(A\mid E)$ becomes
\begin{equation}
    H(A\mid E) = \omega_U \cdot 1+\omega_M \cdot 0 = \omega_U 
\end{equation}
This lead to 
\begin{equation}
    I(A; E) = 1 - H(A\mid E) = 1 - \omega_U = \omega_M
\end{equation}
we know that $\omega_M = p_\parallel/(p_\parallel+p_\perp)$, this lead to
Therefore, Eve’s mutual information per sifted detected bit is:
\begin{equation}
    I(A;E) = \frac{p_\parallel}{p_\parallel + p_\perp}
      = \frac{1}{1 + r}
\end{equation}
Combining Eq.~(\ref{eq4}) and (\ref{eq21}), we have $I(A;B) = 1-h_2(e_B = r/(2(1+r)))$. Therefore, Eve could satisfy the condition needed to hack i.e $I(A;E)>I(A;B)$. But, for Eve tobe not detected she also need to reduce the QBER, $e_B<0.11$ (for BBM92)).
\begin{figure}[t]
    \centering
    \includegraphics[width=0.5\textwidth]{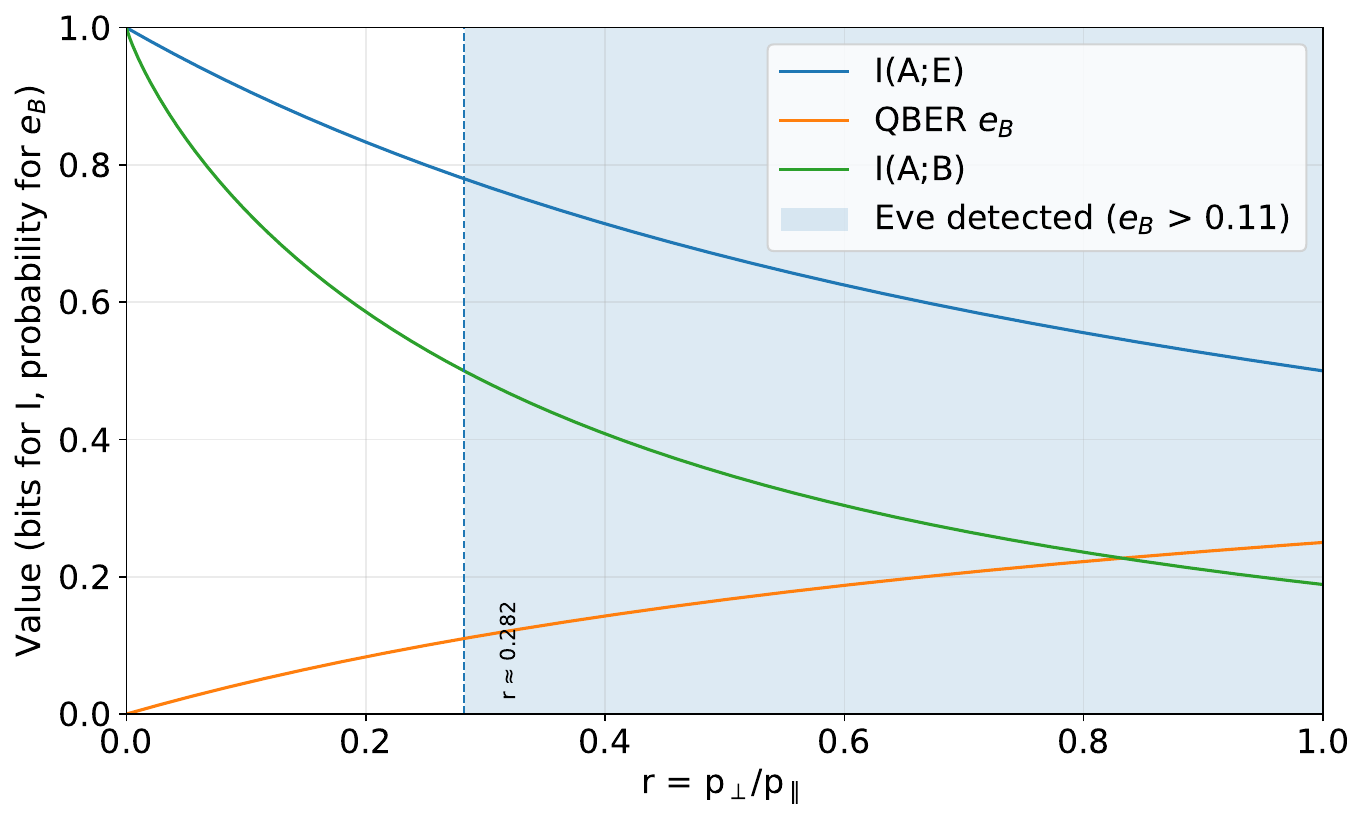}
    \caption{Mutual information $I(A;E)$ (Alice--Eve) and $I(A;B)$ (Alice--Bob) as a function of basis ratio $r$. The shaded region indicates where Eve is detected, assuming detection if $e_B>0.11$. The vertical dashed line is at the threshold $r_{th} = 0.282$.}
    \label{fig2}
\end{figure}
\section{Experiment}\label{exp}
To characterize the dependence of detector dead time on count rate, we injected broadband thermal light (Thorlabs SLS201L/M) into an Excelitas avalanche photodiode (APD, SPCM-AQRH-14-FC). The optical intensity was controlled using calibrated neutral density filters to vary the detected count rate over several orders of magnitude. The detector was operated in free-running mode, and timestamps were recorded using a time-tagging module with 8 ps resolution (We used swabian made time tagger ultra). The effective dead time was determined from the onset of the first non-zero bin in the inter-arrival-time histogram.

The effective dead time was extracted from the histogram of adjacent photon arrival time differences, following the method described in Ref. \cite{kuniyil2025optimizing}. As the incident photon flux increases, the probability of closely spaced detection events rises; however, events occurring within the detector recovery interval are suppressed due to dead time. Consequently, the short-time region of the inter-arrival-time histogram exhibits a depletion whose width corresponds to the effective recovery time of the detector. The minimum observable separation between consecutive detection events therefore provides a direct estimate of the dead time.

By repeating this measurement at different count rates, we obtained the count-rate–dependent effective dead time $t_d(\lambda)$.
\begin{figure*}[t]
    \centering
    \includegraphics[width=\textwidth]{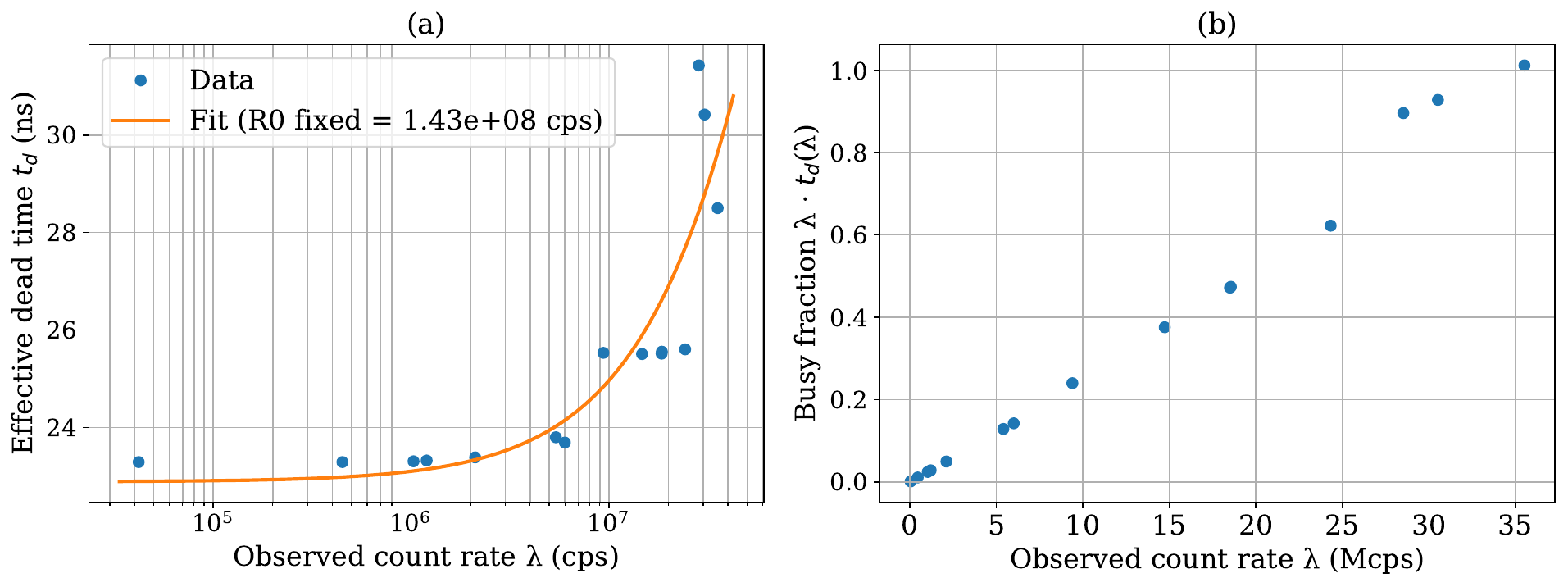}
    \caption{The response of detector SPCM-AQRH-14-FC in various count rate. (a) Measured effective detector dead time $t_d$ as a function of the observed count rate. The solid curve shows a fit to the recovery model presented in Ref~\cite{kuniyil2025optimizing}. (b) Detector busy fraction $\lambda t_d(\lambda)$ as a function of the observed count rate $\lambda$, calculated from the measured dead time data. The busy fraction represents the proportion of time the detector is unavailable due to recovery. }
    \label{fig3}
\end{figure*}
Fig~\ref{fig3}(a) shows the measured dependence of the effective detector dead time on the observed count rate. The dead time increases from its nominal manufacturer-specified value of 23.3 ns to approximately 31.5 ns at high count rates. In the low-noise regime (count rate < 4 Mcps), the measured dead time remains near 23–24 ns. However, as the count rate exceeds 4 Mcps, the dead time begins to increase and surpasses 30 ns above approximately 25 Mcps. This behavior indicates a clear count-rate–dependent recovery nonlinearity in the detector.

Since the recovery-induced erasure model relies on temporary detector unavailability during the dead-time interval, it is necessary to estimate the fraction of time the detector is insensitive. In the low-count regime, where $t_d = 23.3$ ns and the count rate is small, the detector is effectively available most of the time. To quantify the fraction of time the detector is unavailable, we introduce the dimensionless parameter
\begin{equation}
    \lambda t_d(\lambda)
\end{equation}
referred to as the busy fraction. The busy fraction represents the proportion of time the detector resides within its recovery (dead-time) window. As the count rate increases, the busy fraction rises rapidly and exceeds 0.9 above approximately 25 Mcps. The experimentally estimated busy fraction is shown in Fig.~\ref{fig3}(b).

\begin{figure}[t]
\centering
\includegraphics[width=0.5\textwidth]{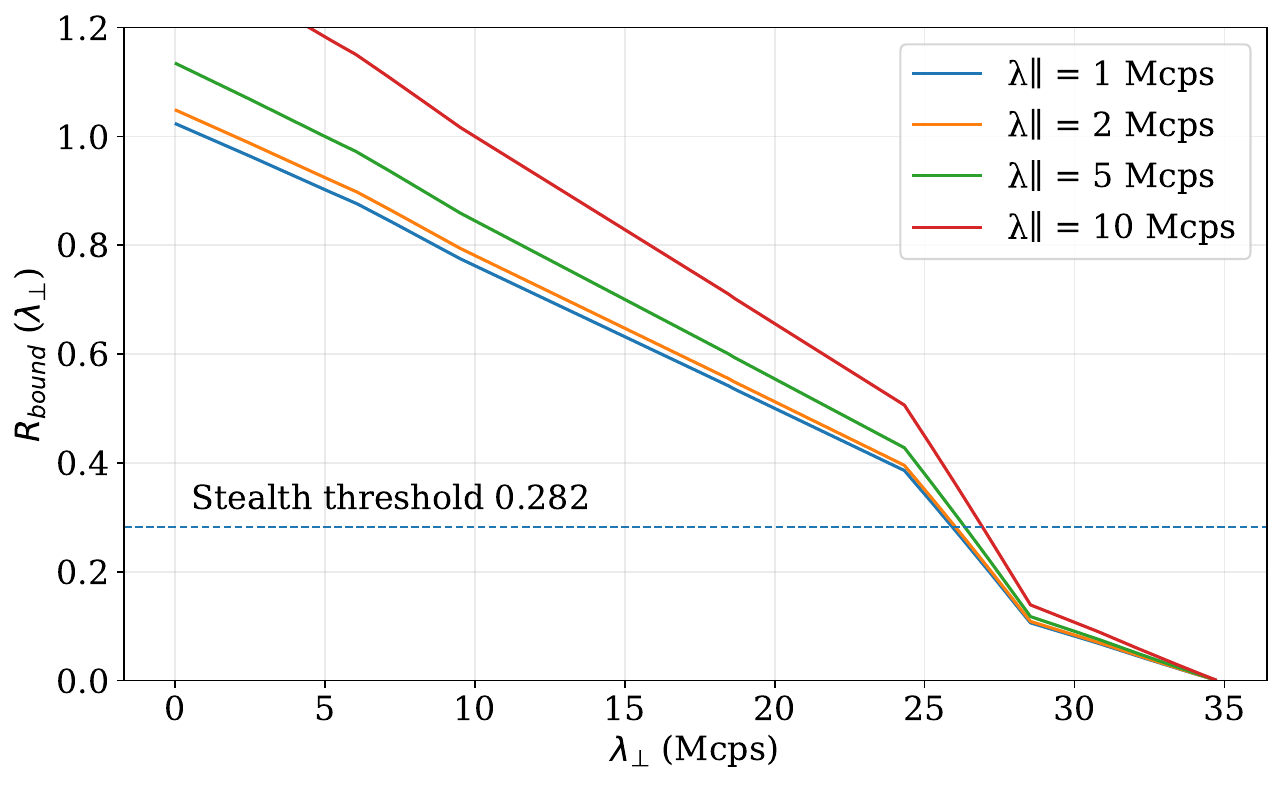}
\captionsetup{width=.9\linewidth}
\caption{Conservative bound on the ratio $R_{\text{bound}}$ as a function of the orthogonal-basis loading rate $\lambda_\perp$ \ computed using the measured dead-time dependence $t_d(\lambda)$.}
\label{fig4}
\end{figure}

To assess the feasibility of the Recovery-Induced Erasure Attack, we next estimate the ratio $r = p_\perp/p_\parallel$. Assuming Poisson arrival statistics for the noise photons and adopting a conservative non-paralyzable detector model, the signal detection probability under loading satisfies
\begin{equation}
    p(\lambda)\leq
p_0(1-\lambda t_d(\lambda)).
\end{equation}
Combining Eqs.~(\ref{eq11}) and (\ref{eq12}), this yield a conservative bound: 
\begin{equation}
    R_{\text{bound}}(\lambda_\perp) = \frac{1 - \lambda_\parallel \, t_d(\lambda_\parallel)}{1 - \lambda_\perp \, t_d(\lambda_\perp)}.
\end{equation}
For representative aligned-basis loading rates $\lambda_\parallel= $ 1, 2, 5, 10 Mcps we evaluate this bound using interpolation of the measured $t_d$. The resulting stealth regime is shown in Fig.~\ref{fig4}. The dashed horizontal line at 0.282 corresponds to the QBER abort threshold of $11\%$. Curves falling below this threshold indicate parameter regimes in which QBER suppression is achievable under the conservative availability model. Notably, this condition is satisfied only when $\lambda_\perp$ enters the high-count regime (upper tens of Mcps), where detector recovery nonlinearity becomes pronounced.
\section{discussion}\label{discussion}
The RIE attack works when the dead time widening happen as a result of increased count rate. We tested the attacking with weaker Eve, non-deterministic pre-pulse model that provides a conservative bound. In the presence of deterministic pre-pulse timing, suppression of the signal detector can be stronger than predicted by the average busy-fraction model. In our attacking test, we used thermal light as a pre-pulse model which has random (Poisson) arrivals, steady-state statistics, and has no precise timing control. Therefore, it gives an average availability.
If Eve sends a strong pre-pulse at time $t=0$ and a signal pulse at time $t =\Delta$, then detector behavior is not statistical — it is time conditioned.

For a non-paralyzable detector, detector is unavailable if $\Delta<t_d$. Therefore, we can assume, in an Eve with deterministic pre-pulse, we have 
\begin{equation}
    p_{\text{click}} = \begin{cases}
0, & \Delta < t_d \\
p_0, & \Delta > t_d.
\end{cases}
\end{equation}
This is a step-like suppression. Therefore, Eve does not require to elevate the count rate as in non-deterministic pre-pulse. Under deterministic pre-pulse model the suppression of desirable basis is maximum. Meaning, with deterministic model Eve can achieve $p_\perp\rightarrow 0$ while keeping $p_\parallel\rightarrow p_0$. If the pre-pulse is routed to the non-signal detector in the aligned case, this yields \(r = 0\), which is much lower than in the case where a non-deterministic pulse is used as a pre-pulse, for which \(r = 0.282\).

Important distinction of RIE attack with well known attacks are counter measure based on measurement window enlargement does not help to deviate this attack. In detector-efficiency-mismatch and time-shift attacks, enlarging the coincidence window is a common countermeasure because it restores symmetry in timing-dependent detection efficiencies. In such attacks, detection events are present but temporally biased, and a wider coincidence window ensures that shifted clicks are not preferentially accepted or rejected. In contrast, the recovery-induced erasure mechanism analyzed here operates at the level of detector availability rather than timing discrimination. When a detector is within its recovery (dead-time) interval at the photon arrival time, no detection event is generated and consequently no timestamp exists to be evaluated by the coincidence logic. Therefore, increasing the coincidence window cannot recover suppressed detection events, as erasures correspond to physically missing clicks rather than timing shifts. Indeed, enlarging the coincidence window may increase the accidental coincidence rate, potentially raising the observed QBER without mitigating the underlying erasure asymmetry. This distinction highlights that countermeasures designed for efficiency-mismatch attacks do not directly address recovery-induced erasure mechanisms.

Although the recovery-induced erasure mechanism has been formulated here in the context of active-basis BBM92 implementations, the underlying principle is not restricted to this specific architecture. The essential requirement of the attack is the presence of count-rate–dependent detector recovery nonlinearity that induces basis-dependent detection suppression $p_{\perp} < p_{\parallel} $. This condition does not rely on entanglement-based key generation per se, but rather on the structure of the measurement stage and detector behavior.

In passive-basis BBM92 receivers, where basis selection is implemented using a beamsplitter rather than an active modulator, the attack mechanism remains applicable provided that differential loading of detector branches can be achieved. Since passive receivers inherently split the optical power between basis arms, pre-pulse–induced loading may distribute across multiple detectors, potentially reducing asymmetry. However, if recovery-induced suppression can be biased toward specific detection channels—either via polarization-selective illumination or differential coupling—the same erasure asymmetry can arise. The effectiveness in passive systems therefore depends on the degree of controllable detector loading rather than the basis selection mechanism itself.

For prepare-and-measure protocols such as BB84, the recovery-induced erasure mechanism is similarly applicable at the receiver stage. In active-basis BB84 implementations, where detectors corresponding to different bases are separated spatially or temporally, count-rate–dependent recovery can again induce asymmetric detection probabilities. The resulting erasure channel suppresses mismatch events while preserving aligned-basis detections, leading to QBER reduction accompanied by increased loss. In passive BB84 receivers employing a beamsplitter for basis choice, the symmetry of optical splitting may reduce the magnitude of achievable asymmetry, but does not fundamentally eliminate the recovery-based suppression effect.

More generally, the attack does not depend on whether key bits are derived from coincidence measurements (BBM92) or single-detector events (BB84), but rather on the existence of independent detectors whose recovery dynamics can be differentially loaded. Therefore, any QKD architecture employing free-running avalanche photodiodes with count-rate–dependent dead time is, in principle, susceptible to recovery-induced erasure effects, although the achievable stealth regime and required loading levels will depend on the specific receiver configuration.

The recovery-induced erasure mechanism discussed above relies on inducing count-rate–dependent suppression in trusted detection units at the receiver. In standard BB84 and BBM92 implementations, the detectors are located within the legitimate parties’ measurement stations and are implicitly trusted in the security model. Consequently, recovery nonlinearity can translate directly into basis-dependent erasure asymmetry and altered key statistics.

In contrast, measurement-device-independent QKD (MDI-QKD) fundamentally alters the threat model by relocating the measurement apparatus to an untrusted relay performing Bell-state measurements \cite{lo2012measurement}. In MDI-QKD, detection events at the measurement station are not assumed to be trustworthy, and security is derived from the time-reversed entanglement structure rather than detector behavior. Therefore, recovery-induced suppression at the untrusted measurement node does not directly yield the same information advantage to an external eavesdropper, as detector manipulation is already incorporated into the adversarial model.

However, practical MDI-QKD implementations may still be indirectly affected if recovery nonlinearity introduces basis-dependent detection efficiencies that violate assumptions used in parameter estimation or decoy-state analysis. In particular, if count-rate–dependent dead time alters the relative yield of specific Bell-state projections in a basis-correlated manner, it could modify the observed gains and error rates. While such effects do not compromise the fundamental security proof of MDI-QKD—which is detector-independent—they may impact finite-key performance, parameter estimation accuracy, or system stability.

Thus, recovery-induced erasure does not constitute a direct security vulnerability in ideal MDI-QKD in the same manner as in trusted-detector BB84 or BBM92 systems. Nevertheless, detector recovery nonlinearity remains relevant for implementation robustness and performance analysis in practical MDI-QKD deployments.

The recovery-induced erasure mechanism exploits count-rate–dependent detector nonlinearity to convert mismatch-induced errors into loss while preserving low QBER. Effective countermeasures therefore require monitoring and constraining detector operating conditions rather than solely relying on coincidence-window adjustments. A primary mitigation strategy is continuous monitoring of detector singles rates and recovery statistics. Real-time surveillance of count rates and inter-arrival-time distributions can reveal abnormal loading conditions or shifts in effective dead time. Such monitoring approaches have been proposed and experimentally demonstrated in response to bright-light and detector-control attacks, where abnormal photocurrent or count-rate signatures were used to detect adversarial illumination \cite{lydersen2010hacking, yuan2011resilience}.

Optical power monitoring at the receiver input, using calibrated taps and classical photodiodes, has likewise been suggested to detect injected bright pulses or anomalous background illumination \cite{jain2011device}. In addition, enforcing strict upper bounds on allowable count rates and aborting operation when deviations from expected detector behavior are observed can mitigate recovery-based manipulation. From a hardware perspective, detector multiplexing (employing parallel SPADs per logical detector channel) reduces vulnerability to dead-time saturation by lowering the effective busy fraction.

\section{Conclusion}
This work establishes count-rate–dependent detector recovery as a security-relevant vulnerability in practical QKD systems. We developed a Recovery-Induced Erasure (RIE) attack model showing how variations in SPAD dead time under high count-rate loading can be exploited to engineer basis-dependent detection suppression in active BBM92 implementations. By converting mismatch-induced errors into losses, the attack reduces the observed QBER while increasing erasure probability, thereby enabling a stealth regime below the protocol abort threshold. Importantly, we demonstrated that this regime can be achieved even under a conservative adversarial model in which Eve is limited to a weak, non-deterministic pre-pulse source.

Our analysis quantitatively identifies the operational conditions under which recovery-induced erasure becomes effective and highlights the role of detector recovery nonlinearity in shaping security-relevant statistics. These results show that detector dead time cannot be treated as a fixed, benign parameter in practical QKD security analyses. Finally, we discussed mitigation strategies based on detector monitoring and recovery-aware countermeasures, emphasizing the need to explicitly incorporate detector recovery dynamics into implementation-level security models.
\bibliography{ref}
\bibliographystyle{IEEEtran}
\end{document}